\documentclass[aps,prl,amsmath,amssymb,nofootinbib,reprint,longbibliography,linenumbrs,superscriptaddress]{revtex4-1}
\pdfoutput=1
\usepackage{graphicx}
\usepackage{dcolumn}
\usepackage{bm}

\usepackage[T1]{fontenc} 
\usepackage{amsmath}
\usepackage{amssymb}
\usepackage{amsfonts}
\usepackage{mathtools}
\usepackage{microtype}
\usepackage{tikz}
\usepackage{bm}
\usepackage[utf8]{inputenc}
\usepackage{blkarray}
\usepackage{bigstrut}
\usepackage{nccmath}
\usepackage{enumitem}
\usepackage{braket}
\usepackage{dsfont}
\usepackage{lipsum}
\usepackage{physics}
\usepackage{xcolor}
\usepackage{color}
\usepackage{float}
\usepackage{subfig}
\usepackage{hyperref}

\usepackage[export]{adjustbox}

\newcommand{\masters}{\nu}
\newcommand{\lbl}{Loop-by-Loop~}

\newcommand{\overbar}[1]{\mkern 1.5mu\overline{\mkern-1.5mu#1\mkern-1.5mu}\mkern 1.5mu}

\definecolor{hjaltegreen}{rgb}{0.0,0.5,0.0}

\begin{document}


\title{Vector Space of Feynman Integrals and Multivariate Intersection Numbers}

\author{Hjalte Frellesvig}
 \email{hjalte.frellesvig@pd.infn.it}
\author{Federico Gasparotto}
 \email{federico.gasparotto@pd.infn.it}
\author{Manoj~K.~Mandal}
 \email{manojkumar.mandal@pd.infn.it}
\author{Pierpaolo Mastrolia}
 \email{pierpaolo.mastrolia@pd.infn.it}

\affiliation{Dipartimento di Fisica e Astronomia, Universit\`a di Padova, Via Marzolo 8, 35131 Padova, Italy}
\affiliation{INFN, Sezione di Padova, Via Marzolo 8, 35131 Padova, Italy}

\author{Luca Mattiazzi}
 \email{luca.mattiazzi@pd.infn.it}
 
\affiliation{INFN, Sezione di Padova, Via Marzolo 8, 35131 Padova, Italy}
\affiliation{Dipartimento di Fisica e Astronomia, Universit\`a di Padova, Via Marzolo 8, 35131 Padova, Italy}

\author{Sebastian Mizera}
 \email{smizera@pitp.ca}

\affiliation{Perimeter Institute for Theoretical Physics, Waterloo, ON N2L 2Y5, Canada}
\affiliation{Department of Physics \& Astronomy, University of Waterloo, Waterloo, ON N2L 3G1, Canada}

\date{\today}

\begin{abstract}
Feynman integrals obey linear relations governed by intersection numbers, which act as scalar products between vector spaces.
We present a general algorithm for constructing multivariate intersection numbers relevant to Feynman integrals, and show for the first time how they can be used to solve the problem of integral reduction to a basis of master integrals by projections, and to directly derive functional equations fulfilled by the latter. 
We apply it to the derivation of contiguity relations for special functions admitting multi-fold integral representations, and to the decomposition of 
a few Feynman integrals at one- and two-loops,
as first steps towards potential applications to generic multi-loop integrals. 
\end{abstract}

\maketitle


\section{Introduction}

Scattering amplitudes encode crucial information about collision phenomena in our universe, from the smallest to the largest scales. 
Within the perturbative field-theoretical approach, the evaluation of multi-loop Feynman integrals is a mandatory operation for the determination of scattering amplitudes and related quantities.

Linear relations among Feynman integrals can be exploited to simplify the evaluation of scattering amplitudes: they can be used both for decomposing scattering amplitudes in terms of a basis of functions, referred to as {\it master integrals} (MIs), and for the evaluation of the latter. 
The standard procedure used to derive relations among Feynman integrals in dimensional regularization makes use of {\it integration-by-parts identities} (IBPs) \cite{Chetyrkin:1981qh}, which are found by solving linear systems of equations \cite{Laporta:2001dd} (see \cite{Grozin:2011mt,Zhang:2016kfo} and references therein for reviews). Algebraic manipulations in these cases are very demanding, and 
efficient algorithms for solving large-size systems of linear equations have been recently devised, by making use of finite field arithmetic and rational functions reconstruction \cite{vonManteuffel:2014ixa,Peraro:2016wsq,Peraro:2019svx}.

In \cite{Mastrolia:2018uzb}, it was shown that {\it intersection numbers} \cite{cho1995} of differential forms can be employed to define (what amounts to) a {\it scalar product} on a \emph{vector space} of Feynman integrals in a given family.
Using this approach, projecting any multi-loop integral onto a basis of MIs is conceptually no different from decomposing a generic vector into a basis of a vector space. Within this new approach, relations among Feynman integrals can be derived avoiding the generation of intermediate, auxiliary expressions which are needed when applying Gauss elimination, as in the standard IBP-based approaches.

In the initial studies, \cite{Mastrolia:2018uzb,Frellesvig:2019kgj}, this novel decomposition method was applied to the realm of special mathematical functions falling in the class of Lauricella functions, as well as to Feynman integrals on maximal cuts, {\it i.e.} with on-shell internal lines, mostly admitting a one-fold integral representation. 
Those results concerned a partial construction of Feynman integral relations, mainly limited to the determination of the coefficients of the MIs with the same number of denominators as the integral to decompose, which was achieved by means of intersection numbers for {\it univariate} forms. 

In this paper, we make an important step further, and address the {\it complete} integral reduction, for the determination of {\it all} coefficients, including those associated to MIs corresponding to sub-graphs. 
In the current work, we discuss the one-loop massless four-point integral as a paradigmatic case, although the algorithm has been successfully applied to several other cases at one- and two-loop.

Generic Feynman integrals admit multi-fold integral representations. Their complete decomposition requires the evaluation of intersection numbers for multivariate rational differential forms. 
Intersection numbers of multivariate forms have been previously studied in \cite{matsumoto1994,matsumoto1998,OST2003,doi:10.1142/S0129167X13500948,goto2015,goto2015b,Yoshiaki-GOTO2015203,Mizera:2017rqa,2019arXiv190401253M}.
Recently, a new recursive algorithm was introduced in \cite{Mizera:2019gea}.
In this letter, we present its refined implementation and application to Feynman integrals, which provide a major step towards large-scale applicability of our strategy for the reduction to MIs.
The results of this work show potential for further applications ranging
from particle physics, through condensed matter and statistical mechanics, to gravitational-wave physics, while making new connections to mathematics.

\section{Integrals and Differential Forms}
In this work, we focus on integrals of the hypergeometric type,
\begin{eqnarray}
I = \int_{{\cal C}} u({\bf z}) \, \varphi(\bf{z}) \ ,
\label{eq:def:I:basic}
\end{eqnarray}
where: 
${\bf z} = ({z_1,\ldots,z_n})$ are integration variables;
${\cal C}$ is the integration domain;
$u$ is a multi-valued function of the form $u = \prod_i \mathcal{B}_i(\mathbf{z})^{\gamma_i}$
with $\gamma_i \notin {\mathbb Z}$,   
such that $\prod_i \mathcal{B}_i$ vanishes on the integration boundary $\partial{{\cal C}}$;
and $\varphi$ is a single-valued differential $n$-form, 
\begin{eqnarray}
\varphi({\bf z}) &=& {\hat \varphi}({\bf z}) \, d^n {\bf z}  \ , \qquad 
d^n{\bf z} \equiv d z_1 \wedge \ldots \wedge d z_n \ ,
\label{eq:def:phi:generic}
\end{eqnarray}
with ${\hat \varphi}$ being a rational function with all poles regulated by $u(\mathbf{z})$.
Then employing Stokes' theorem we find equivalence classes of $n$-forms,
\begin{equation}
\label{eq:equivalence-relation}
    \varphi \sim \varphi + \nabla_\omega \xi,
\end{equation}
for any $(n{-}1)$-form $\xi$ and
where $\nabla_\omega \equiv d + \omega\wedge$ is a covariant
derivative with a one-form $\omega \equiv d\log u$. 
The space of $n$-forms modulo the relation eq.~\eqref{eq:equivalence-relation} forms a vector space called a \emph{twisted cohomology group}\footnote{%
We refer the interested reader to \cite{aomoto2011theory,yoshida2013hypergeometric,Mizera:2019gea} for reviews of twisted (co)homologies and their intersection theory, as well as \cite{Mizera:2017rqa,Mizera:2017cqs,Mastrolia:2018uzb,Frellesvig:2019kgj,Mimachi2003,Mimachi2004,Frost:2018djd,Li:2018mnq,Brown:2018omk,Mizera:2019gea} and \cite{delaCruz:2017zqr,Abreu:2017enx,Abreu:2018nzy,delaCruz:2019skx}
for some recent applications of these ideas to physics.
} $H_{\omega}^n$. We denote its elements by $\langle \varphi| \in H^n_\omega$. Within this framework,
the integral $I$ from eq.~\eqref{eq:def:I:basic} can be interpreted as a pairing of $\langle \varphi |$ with the integration contour $| {\cal C}]$, 
\begin{eqnarray}
I = \langle \varphi|{\cal C}] \ .
\label{eq:def:I:pairing}
\end{eqnarray}
Since in our applications $|\mathcal{C}]$ will always stay constant, the vector space of such integrals is the same as that of $\langle \varphi |$.

Consider a set of $\nu$ MIs, say $J_i$, defined as 
\begin{eqnarray}
J_i = \int_{\cal C} u({\bf z}) \, e_i({\bf z}) 
    = \langle e_i |{\cal C}]
\ , \qquad 
i=1,\ldots,\nu \, ,
\end{eqnarray}
in terms of any independent set of differential forms $\langle e_i|$. 
Then, the decomposition of a generic integral $I$ in terms of the MIs $J_i$,
\begin{eqnarray}
I = \sum_{i=1}^\nu c_i \, J_i \ ,
\label{eq:I:deco}
\end{eqnarray}
can be interpreted as coming from the more fundamental decomposition of the differential form $\langle \varphi|$ in terms of the basis forms $\langle e_i|\,$,
namely 
\begin{eqnarray}
\langle \varphi | = \sum_{i=1}^\nu c_i \, \langle e_i| \ ,
\end{eqnarray}
with the coefficients determined by the {\it master decomposition formula} \cite{Mastrolia:2018uzb,Frellesvig:2019kgj},
\begin{eqnarray}
c_i = \sum_{j=1}^\nu \langle \varphi | h_j \rangle \big({\mathbf C}^{-1}\big)_{ji} \ , 
\qquad
{\mathbf C}_{ij} = \langle e_i | h_j \rangle\, \ ,
\label{eq:masterdeco}
\end{eqnarray}
where 
$| h_j \rangle$ ($j=1,\ldots,\nu$)%
\footnote{%
Suitable choices of the basis forms 
$\langle e_i|$ and $|h_i \rangle$ can be made,
such that ${\mathbf C} = {\mathbb I}_{\nu \times \nu}$, 
yielding a simplified decomposition formula \cite{Mastrolia:2018uzb,Frellesvig:2019kgj}, 
$\langle \varphi | = \sum_{i=1}^\nu \langle \varphi | h_i \rangle \, \langle e_i|$.
}, span a {\it dual} (and auxiliary) vector space $(H^{n}_\omega)^\ast = H^{n}_{-\omega}$. The scalar product $\langle \varphi_L | \varphi_R \rangle$ between the two vector spaces is called an \emph{intersection number} of differential forms \cite{cho1995}.
 
Using eqs.~(\ref{eq:I:deco},\ref{eq:masterdeco}), our algorithm for expressing any 
integral of the type of eq.~\eqref{eq:def:I:basic} as linear combinations of MIs proceeds along three steps: 
\begin{enumerate}[leftmargin=1.2em]
\item Determination of the number $\nu$ of MIs.
\item Choice of the bases of forms $\langle e_i|$ and $|h_i \rangle$. 
\item Evaluation of the intersection numbers for multi-variate forms, 
appearing in the entries of the ${\mathbf C}$-matrix,
and in $\langle \varphi | h_j \rangle$.
\end{enumerate}

\noindent
{\bf Number of Master Integrals.} Under some assumptions one can show that all other vector spaces $H^{k \neq n}_{\pm \omega}$ are trivial, which means that $\varphi$ can only be $n$-forms \cite{aomoto1975vanishing}. In those cases the dimension of these vector spaces, {\it i.e.} the number $\masters$ of MIs, can be determined topologically%
\footnote{In the Feynman integral literature, the finiteness of $\masters$ was first considered in \cite{Smirnov:2010hn}, while its connection to 
the number of critical points and Euler characteristics were previously explored in \cite{Lee:2013hzt,Aluffi:2008sy,Marcolli:2008vr,Bitoun:2017nre,Mastrolia:2018uzb,Frellesvig:2019kgj}.},
\begin{equation}
\masters \equiv \dim H^{n}_{\pm \omega} = (-1)^{n} \left( n{+}1 - \chi({\cal P}_\omega) \right) \ ,
\end{equation}
in terms of the Euler characteristic $\chi({\cal P}_\omega)$ of the projective variety $\mathcal{P}_\omega$ defined as the set of poles of $\omega$.

This connection allows us to use complex Morse (Picard--Lefschetz) theory to determine $\nu$ as the number of critical points of the function $\log u(\mathbf{z})$. Let us define
\begin{eqnarray}
\omega &\equiv& d {\log}\, u(\mathbf{z}) = \sum_{i=1}^n {\hat \omega}_i \, dz_i \ ,
\label{eq:omegadef}
\end{eqnarray} 
then the number of critical points is given by the number of solutions of 
the system of equations 
\begin{eqnarray}
{\hat \omega}_i \equiv {\partial_{z_i} \!\log u(\mathbf{z})} =  0\,, \qquad i=1,\ldots,n \ ,
\label{eq:criticalpoints}
\end{eqnarray}
with the short-hand notation $\partial_{z_i} \equiv \partial/\partial z_i$, provided that the set of solutions is finite.
Additional details are provided in the App. A.

\section{Intersection Numbers}

In this section we review a recursive algorithm for the evaluation of intersection numbers of multivariate differential forms introduced in \cite{Mizera:2019gea}.

We start by decomposing the $n$-dimensional space with coordinates $(z_1,\ldots,z_n)$ 
into a $(n{-}1)$-dimensional subspace parametrized by $(z_1,\ldots,z_{n-1})$, which we call {\it inner} space, and a one-dimensional subspace with $z_n$, dubbed {\it outer} space. The aim is to express the original intersection number $_{\bf n}\langle \varphi_L^{({\bf n})} | \varphi_R^{(\bf n)} \rangle$ in terms of one-dimensional residues on the outer space and intersection numbers ${}_{\mathbf{n-1}}\langle \ldots | \ldots \rangle$ on the inner space, which are assumed to be known at this stage. 
The choice of the variables (and their ordering) parametrizing the inner and outer spaces is arbitrary: in the following, we use the generic notation $\mathbf{m} \equiv (12\dots m)$ to denote the variables taking part in a specific computation.

Thus, 
the original ${\bf n}$-forms can be decomposed according to 
 \begin{eqnarray}
 \label{varphi-L-projection}
\langle \varphi_{L}^{({\bf n})} |
 &=& \sum_{i=1}^{\nu_{\bf{n-1}}} \langle e^{{\bf (n-1)}}_i | \wedge 
 \langle \varphi_{L,i}^{(n)} | \ , \\ 
 | \varphi_{R}^{({\bf n})} \rangle &=& 
 \sum_{i=1}^{\nu_{\bf{n-1}}} |h^{{\bf (n-1)}}_i \rangle \wedge 
 | \varphi_{R,i}^{(n)} \rangle 
 \ ,
 \end{eqnarray}
where $\nu_{\bf{n-1}}$ is the number of master integrals on the inner space with arbitrary bases $\langle e^{{\bf (n-1)}}_i |$, $|h^{{\bf (n-1)}}_j \rangle$ and the metric matrix
\begin{equation}
\big({\mathbf C}_{{\bf (n-1)}}\big)_{ij} \equiv {}_{\bf{n-1}}\langle e_i^{({\bf n-1})} | h_j^{({\bf n-1})} \rangle \ .
\end{equation}
In the above expressions $\langle \varphi_{L,i}^{(n)} |$ and $| \varphi_{R,j}^{(n)} \rangle$ are $dz_n$-forms treated as coefficients of the basis expansion. They can be obtained by a projection similar to eq.~\eqref{eq:masterdeco}, for example:
\begin{equation}\label{eq:varphi-R}
| \varphi_{R,i}^{(n)} \rangle = \big({\mathbf C}_{({\bf n-1})}^{-1}\big)_{ij}\;
{}_{\bf{n-1}}\langle e_j^{({\bf n-1})} | \varphi_R^{(\mathbf{n})} \rangle \ ,
\end{equation}
where from now on we use the implicit sum notation for repeated indices.
The recursive formula for the intersection number reads
\begin{eqnarray}
\hspace*{-0.5cm}
{}_{\bf n}\langle \varphi_L^{({\bf n})} | 
\varphi_R^{({\bf n})} \rangle 
{=} -\!\!\! \sum_{p \in {\cal P}_{n} } \!
\underset{z_n = p}{\Res} \!
\Big( 
{}_{\bf n-1}\langle \varphi_L^{(\mathbf{n})} | h_i^{(\mathbf{n-1})} \rangle\,
\psi^{(n)}_i 
\Big) \ ,
\label{eq:multivarIntNumb}
\end{eqnarray}
where functions $\psi^{(n)}_i$ are solutions of the system of differential equations
\begin{eqnarray}
\partial_{z_n} 
  \psi^{(n)}_i 
- {\hat{\mathbf\Omega}}^{(n)}_{ij} \psi^{(n)}_j  = 
{\hat \varphi}^{(n)}_{R,i} \ ,
\label{eq:sysofdeq}
\end{eqnarray}
where $\langle \varphi_{R,i}^{(n)} | = \hat\varphi_{R,i}^{(n)} dz_n$ from eq.~\eqref{eq:varphi-R}.
The $\nu_{\mathbf{n-1}}{\times}\nu_{\mathbf{n-1}}$ matrix ${\hat{\mathbf\Omega}}^{(n)}$
given by
\begin{eqnarray}
\label{Omega-n-2}
{\hat{\mathbf\Omega}}^{(n)}_{ij} = -\big({\mathbf C}_{({\bf n-1})}^{-1}\big)_{ik}\;
{}_{\bf{n-1}}\langle 
e^{{\bf (n-1)}}_k | (\partial_{z_n} {-} \hat \omega_{n}) h^{{\bf (n-1)}}_j
\rangle ,\;\;
\end{eqnarray}
and finally ${\cal P}_{n}$ is the set of poles of ${\hat{\mathbf\Omega}}^{(n)}$ given by the union
of the poles of its entries (including possible poles at infinity).

We observe that the solution of eq.~\eqref{eq:sysofdeq} around $z_n{=}p$ can be formally written 
in terms of a path-ordered matrix exponential
\begin{eqnarray}
\vec{\psi}^{(n)} {=} \left( \int^{z_n}_{p} \!\!\vec{\varphi}^{(n)}_{R}\!(y)\, {\cal P} e^{-\!\!\int^y_p \!{\mathbf\Omega}^{(n)}(w)} \right)\!\!\left( {\cal P} e^{\!\int^{z_n}_p \!{{\mathbf\Omega}}^{(n)}(w)} \right)\!\;\;\;
\end{eqnarray}
for a vector $\vec{\psi}^{(n)}$ with entries $\psi^{(n)}_i$. Nevertheless for its use in eq.~\eqref{eq:multivarIntNumb}, it is sufficient to know only a few leading orders of ${\vec{\psi}^{(n)}}$ around each $p \in {\cal P}_{n}$. 
Therefore, it is easier to find the solution of the system eq.~\eqref{eq:sysofdeq} by a {\it holomorphic} Laurent series expansion, using an ansatz for each component $\psi^{(n)}_i$, see \cite{Mastrolia:2018uzb,Frellesvig:2019kgj}. Such a solution exists if the matrix $\Res_{z_n = p} {\mathbf\Omega}^{(n)}$ does not have any non-negative integer eigenvalues, which we assume from now on.

The recursion terminates when $n{=}1$, in which case the inner space is trivial: $\nu_{\mathbf{0}} = \langle e_1^{(\mathbf{0})}| = | h_1^{(\mathbf{0})}\rangle = 1$, and we impose the initial conditions
\begin{eqnarray}
{\hat{\mathbf\Omega}}^{(1)}_{11} = \hat{\omega}_1 \, , \quad {}_{\mathbf{0}}\langle \varphi_L^{({\bf 1})} | h_1^{(\bf{0})} \rangle = \varphi_L^{(\bf{n})} ,\quad
\varphi_{R,1}^{(1)} = \varphi_R^{(\bf{n})} .
\end{eqnarray}
In this case eqs.~(\ref{eq:multivarIntNumb},\ref{eq:sysofdeq}) reduce to a computation of univariate intersection numbers \cite{cho1995,matsumoto1998} previously studied in \cite{Mastrolia:2018uzb,Frellesvig:2019kgj}. Plugging everything together, eq.~\eqref{eq:multivarIntNumb} can be expressed as
\begin{align}
{}_{\bf{n}}\langle \varphi_L^{({\bf n})} 
|\varphi_R^{({\bf n})} \rangle 
= &
\,(-1)^n\!\!\! \sum_{p_{n} \in {\cal P}_{n}} \!\!\cdots\!\!
\sum_{ p_{1} \in {\cal P}_{1}} 
\underset{z_n = p_n}{\Res} \cdots
\underset{z_1 = p_1}{\Res}
\nonumber \\
&
\Big( \varphi_L^{({\bf n})}   
\psi_{1 i_{\bf 1}}^{(1)}  
\psi_{i_{\bf 1} i_{\bf 2}}^{(2)} 
\cdots
\psi_{i_{\bf n-2}i_{\bf n-1}}^{(n-1)}
\psi_{i_{\bf n-1}}^{(n)}
\Big) ,
\label{eq:multivarIntNumb:IterativePsi}
\end{align}
where the ranges of summations are $i_{\bf m} = 1,\dots,\nu_{\bf m}$ and each $\psi^{(m)}_{i_{\bf m-1}i_{\bf m}}$ for ${\bf m}= {\bf 1},\dots,{\bf n{-}1}$ is the solution of
\begin{eqnarray}
\label{eq:psi-m-DE}
\partial_{z_m} \psi^{(m)}_{i_{\bf m-1} i_{\bf m}} 
- {\hat{\mathbf\Omega}}^{(m)}_{i_{\bf m-1} j_{\bf m-1} } \psi^{(m)}_{j_{\bf m-1} i_{\bf m}}
= {\hat h}_{i_{\bf m-1} i_{\bf m}}^{(m)} \ ,
\end{eqnarray}
for all $i_{\bf m}$ with $| h_{i_{\bf m-1} i_{\bf m}}^{(m)} \rangle = \hat{h}_{i_{\bf m-1} i_{\bf m}}^{(m)} dz_m$ coming from the projection:
\begin{eqnarray}
| h^{({\bf m})}_{i_{\bf m}} \rangle = 
| h_{i_{\bf m-1}}^{({\bf m-1})} \rangle \wedge
| h_{i_{\bf m-1} i_{\bf m}}^{(m)} \rangle \ ,
\end{eqnarray}
which is known {\it a priori}, because the bases of all inner spaces are arbitrarily chosen. The matrices $\hat{\mathbf\Omega}^{(m)}$ needed in eq.~\eqref{eq:psi-m-DE} are computed analogously to eq.~\eqref{Omega-n-2}. Notice that all $\psi^{(m)}$ entering eq.~\eqref{eq:multivarIntNumb:IterativePsi} need to be computed only {\it once} for a given family of integrals.

The multivariate intersection number given in eqs.~(\ref{eq:multivarIntNumb}, \ref{eq:multivarIntNumb:IterativePsi}) 
is the key formula used in this letter.
Paired with the master decomposition formula eq.~\eqref{eq:masterdeco}, 
the above recursion for intersection numbers yields an expansion of multi-fold integrals of the form in eqs.~(\ref{eq:def:I:basic},\ref{eq:def:I:pairing}) in terms of MIs.
\section{Hypergeometric Function ${}_3 F_2$}

In order to illustrate application of the above algorithm we start with a more familiar case of contiguity relation for the hypergeometric function ${}_3 F_2$.
Consider the function $\mathcal{H}$, defined as,
\begin{align}
    \mathcal{H} \! \left( \! \begin{smallmatrix} a_1 a_2 a_3 \\ b_1 b_2 \end{smallmatrix} \! ; x \right) 
    &\equiv \beta(a_1,b_1{-}a_1) \beta(a_2,b_2{-}a_2) {}_3 F_2 \! \left( \! \begin{smallmatrix} a_1 a_2 a_3 \\ b_1 b_2 \end{smallmatrix} \! ; x \right) \nonumber \\
    &= \int_{\mathcal{C}} u \; d^2 {\bf z}  
    = \langle 1^{(12)} | {\cal C} ] \ ,
    \label{eq:def:H:1}
\end{align}
where $\beta(a,b) = \Gamma(a)\Gamma(b)/\Gamma(a{+}b)$ is the Euler beta-function,
\begin{align}
u &= (1 {-} z_1 z_2 x)^{-a_3} \prod_{i=1}^2 z_i^{a_i{-}1} (1{-}z_i)^{b_i{-}a_i{-}1}
\ , 
\end{align} 
$d^2 {\bf z} = dz_1 \wedge dz_2$, 
and where $\mathcal{C}$ is the square with $z_i \in [0,1]$.
In this case, $\omega$ is defined through eq.~\eqref{eq:omegadef} with
\begin{align}
\hat{\omega}_i = \frac{a_i {-} 1}{z_i} + \frac{1 {+} a_i {-} b_i}{1 {-} z_i} + \frac{a_3 \, x \, z_{i{+}1}}{1 {-} x \, z_1 z_2},
\end{align}
where $z_{i{+}1}$ should be understood as $z_1$ if $i=2$.
The system $\hat{\omega}_1 = \hat{\omega}_2 = 0$ has three solutions, corresponding to $\masters_{(12)}=3$ MIs. We choose three master forms, 
$\langle {e}_i^{(12)}| \equiv \hat{e}_i^{(12)} d^2{\bf z} $, $(i=1,2,3)$,
\begin{align}
\hat{e}_1^{(12)} 
= \frac{1}{z_1} \,,\quad
\hat{e}_2^{(12)} = \frac{1}{z_2} \,,\quad
\hat{e}_3^{(12)} = \frac{1}{1-z_2} \, , 
\end{align}
which correspond to the following set of MIs,
\begin{align}
\!\!\!
\mathcal{H} \! \big( \! \begin{smallmatrix} a_1{-}1, a_2, a_3 \\ b_1{-}1, b_2 \end{smallmatrix} \! ; x \big)\,, \;
\mathcal{H} \! \big( \! \begin{smallmatrix} a_1, a_2{-}1, a_3 \\ b_1, b_2{-}1 \end{smallmatrix} \! ; x \big)\,, \; 
\mathcal{H} \! \big( \! \begin{smallmatrix} a_1, a_2, a_3 \\ b_1, b_2{-}1 \end{smallmatrix} \! ; x \big)\,. 
\label{eq:hygomasters}
\end{align}
At the same time, we define the dual basis, 
$| {h}_i^{(12)} \rangle \equiv \hat{h}_i^{(12)} d^2{\bf z} $, with
$\hat{h}_i^{(12)} = \hat{e}_i^{(12)}$ $(i=1,2,3)$. 
The decomposition of $\langle{1}| = d^2 \mathbf{z}$ in terms of $\langle {e}_i^{(12)}| $,
\begin{eqnarray}
\langle{1}^{(12)} | = \sum_{i=1}^3 c_i \, \langle {e}_i^{(12)}| \ ,
\end{eqnarray}
yields the decomposition of the function defined in eq.~\eqref{eq:def:H:1} in terms of 
those in eq.~\eqref{eq:hygomasters}, which amounts to a contiguity relation for ${}_3F_2$ functions.
The coefficients $c_i$ are determined by means of eq.~\eqref{eq:masterdeco}, requiring the computation of 12 intersection numbers for two-forms, 
that is 9 elements of the matrix $({\bf C}_{(12)})_{ij} = {}_{(12)}\langle e_i^{(12)} | h_j^{(12)}\rangle$ and 3 entries 
${}_{(12)}\langle 1 | h_j^{(12)} \rangle$ for $i,j=1,2,3$.

To apply formula eq.~\eqref{eq:multivarIntNumb}, we consider the $z_2$-subspace as the inner space. In turn, the number of MIs for the inner space is determined by counting the number of solutions of $\hat{\omega}_2=0$ (w.r.t. $z_2$), giving $\nu_{(2)} = 2$. 
The inner bases are $\langle {e}_i^{(2)}| \equiv \hat{e}_i^{(2)} dz_2$, 
$| {h}_i^{(2)} \rangle \equiv \hat{h}_i^{(2)} dz_2 $
$(i=1,2)$, which we choose to be,
\begin{eqnarray}
\hat{e}_1^{(2)} = \hat{h}_1^{(2)} = \frac{1}{z_2} \,,\quad
\hat{e}_2^{(2)} = \hat{h}_2^{(2)} = \frac{1}{1-z_2}.
\end{eqnarray} 

The individual intersection numbers are too large to be printed here. 
Yet, the final result is rather simple, and, in terms of ${}_3 F_2$-functions, it reads,
\begin{align}
\tilde{c}_0 \, {}_3 F_2 \! \left( \! \begin{smallmatrix} a_1, a_2, a_3 \\ b_1, b_2 \end{smallmatrix} \! ; x \right) 
&= \sum_{i=1}^3 \tilde{c}_i \, {}_3 F_2 \left( X_i \right)
\end{align}
where the $X_i$ are the arguments of the functions ${\cal H}$ in eq.~\eqref{eq:hygomasters} and
\begin{align}
\tilde{c}_0 &= (a_1{-}1) (b_1 {-} b_2) + (a_1 {-} a_2) (b_2 {-} a_3 {-} 1) x \,, \nonumber \\
\tilde{c}_1 &= {(b_1{-}1) (a_1 {-} b_2)} \,, \quad 
\tilde{c}_2 = {(a_2 {-} b_1) (1 {-} b_2)} \,, \\
\tilde{c}_3 &= {(a_1 {-} a_2) (1 {-} b_2) (1 {-} x)} \,. \nonumber
\end{align} 
This relation has been (numerically) verified with Mathematica.
\section{Feynman Integral Decomposition}

Within the Baikov representation (BR), Feynman integrals can be cast in the form\footnote{A prefactor $K$, depending on $d$ and on external kinematic invariants, is dropped while deriving integral relations. However, it is considered for the construction of differential equations and dimensional recurrence relations as discussed in \cite{Frellesvig:2019kgj}.} (\ref{eq:def:I:basic}), 
with either $u(\mathbf{z}) = \mathcal{B}^\gamma(\mathbf{z})$ \cite{Baikov:1996iu,Lee:2010wea} or $u(\mathbf{z}) = \prod_i \mathcal{B}_i(\mathbf{z})^{\gamma_i}$ \cite{Frellesvig:2017aai}.
The factors ${\cal B}$ and ${\cal B}_i$ are graph (Baikov) polynomials, and their exponents depend on the dimensional parameter $d$, 
and on the number of loops and external momenta of the corresponding diagram.
The number $n$ of integration variables corresponds to the number of scalar products formed by the external and the loop momenta. In fact, within BR, the propagators of the diagrams, supplemented by a set of auxiliary propagators (related to the irreducible scalar products), are the integration variables. Therefore,  $\varphi$ in (\ref{eq:def:phi:generic}) can be generically written as
\begin{align}
\varphi({\bf z}) = \hat{\varphi}({\bf z}) d^n{\bf z}  = \frac{f({\bf z}) d^n{\bf z} }{ z_1^{a_1} \cdots z_n^{a_n} } \ ,
\label{eq:def:phi:FeynInt}
\end{align}
where $a_i \in {\mathbb Z}$, and where $f$ is a rational function of ${\bf z}$.
Multiple-cut integrals, identified by the on-shell conditions $z_{i_1} {=} \ldots {=} z_{i_k} {=} 0$, are also of the form (\ref{eq:def:I:basic}), but their integrands depend on fewer integration variables (and their integration contour is modified), see \cite{Mastrolia:2018uzb,Frellesvig:2019kgj}. 

\noindent
{\bf Bottom-up Decomposition.} For a given integral, any set of its denominators identifies a {\it sector}. Therefore, one maximal-cut (when all denominators are cut) corresponds to  each sector.
The number of MIs in each sector can be determined by counting the number of critical points of the corresponding maximal-cut\footnote{When singularities of $\varphi$ are not regulated by $u$, or when $u$ contains factors raised to integer powers, a regulator has to be introduced as discussed in (Sec. 10 and App. A of) ref.~\cite{Frellesvig:2019kgj}.}, using eq.~\eqref{eq:criticalpoints}. 

After determining the number of MIs, the decomposition of Feynman integrals can be obtained by means of eq.~(\ref{eq:masterdeco}). This is done by redefining $u$ by multiplying it with a regulating factor $z_i^{\rho_i}$ for each uncut denominator, where the exponents $\rho_i$ are regulators to be put to zero at the end of the calculation. This is done in order to allow the theory to access sectors where the regulated variables appear as propagators. The determination of coefficients can be performed on unitarity cuts, where the integrands are simpler, and the evaluation of the multivariate intersections requires fewer iterations. A minimal set of {\it spanning cuts} will be sufficient to retrieve the information on the complete decomposition~\cite{Larsen:2015ped}, and then, using the regulated $u$, the master decomposition formula \eqref{eq:masterdeco} yields the  coefficients of those MIs that survive on the cut.
As in the case of IBP-based approaches, additional relations may be obtained from the symmetries of the diagrams, in order to minimize the number of independent integrals.

As discussed in refs. \cite{Mastrolia:2018uzb, Frellesvig:2019kgj} also differential equations in kinematic variables, {\it e.g.} $\partial_s J_i = \sum_j a_{ij} J_j$, can be obtained with the above techniques.
\section{\label{sec:massless-box}Massless Box}
\begin{figure}[H]
    \centering
    \includegraphics[width=0.10\textwidth]{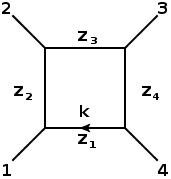}
    \caption{Massless box with massless external legs\\
    ($p_{i}^{2}=0$, for $i=1,2,3,4$). The invariants are  $s=(p_{1}+p_{2})^2$ and $t=(p_{2}+p_{3})^2$.}
    \label{fig:Massles_Box}
\end{figure}

Let us consider the massless box diagram at one loop, Fig.~\ref{fig:Massles_Box}.  
Within the BR, 
\begin{align}
u({\bf z}) 
& = 
\big((st {-} s z_4 {-} t z_3)^2-2 t z_1 (s (t {+} 2 z_3 {-} z_2 {-} z_4) {+} t z_3 ) \nonumber \\
& \!\!\! + s^2 z_2^2 + t^2 z_1^2 - 2 s z_2 (t (s{-}z_3){+}z_4 (s{+}2 t))\big)^{\frac{d-5}{2}}.
\label{eq:def:u:box}
\end{align}

For each of the $15 \, (=2^4{-}1)$ sectors, we use eq.~\eqref{eq:criticalpoints} on the corresponding cut, 
to determine the number $N_{\rm sector}$ of MIs. The non-zero cases are%
\footnote{If the Baikov polynomial $\cal {B}$ is a non-zero constant on the maximal 
cut, the integral is fully localized by the cut-conditions. In this case, the 
condition $\omega = 0$ is always satisfied, and there is $\nu = 1$
master integral.}:
$N_{\{1,2,3,4\}} = 1$, $N_{\{1,3\}} = 1$, $N_{\{2,3\}} = 1$, 
amounting to 3 MIs. We choose them to be:
\begin{align}
& J_1 = \begin{gathered} \includegraphics[width=0.05\textwidth,valign=c]{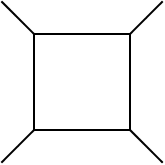} \end{gathered} \, , \quad  J_2= \begin{gathered}  \includegraphics[width=0.05\textwidth,valign=c]{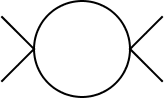} \end{gathered}  \, , \quad J_3= \begin{gathered}  \includegraphics[width=0.03\textwidth,valign=c]{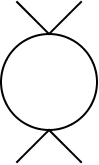} \end{gathered} \, ,
\end{align}
so that any integral $I$ of the form of eq.~\eqref{eq:def:I:basic}, with $u$ given in eq.~\eqref{eq:def:u:box}, and $\varphi$ defined in eq.~\eqref{eq:def:phi:FeynInt} (with $n=4$), can be decomposed as,
\begin{equation}
\begin{gathered} \includegraphics[width=0.05\textwidth,valign=c]{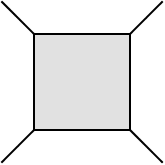} \end{gathered}=c_1\begin{gathered} \includegraphics[width=0.05\textwidth,valign=c]{box_no_label.png} \end{gathered}
+ c_2\, \begin{gathered}  \includegraphics[width=0.05\textwidth,valign=c]{bubble_s_channel.png}\end{gathered}
+c_3\, \begin{gathered}  \includegraphics[width=0.03\textwidth,valign=c]{bubble_t_channel.png} \end{gathered} \ .
\label{eq:masslessbox:decomp}
\end{equation}
We determine the set of spanning cuts as $(\rm Cut_{\{1,3\}},\allowbreak \rm 
Cut_{\{2,4\}})$ to obtain the full decomposition. \\

\noindent
$\bullet$ {$\rm \bf Cut_{\{1,3\}}:$}
On this specific cut, we use the regularized 
$
u_{1,3} = z_2^{\rho_2} z_4^{\rho_4} \,
u(0,z_2,0,z_4)
$
to obtain the corresponding ${\hat \omega}_2$ and ${\hat \omega}_4$. 
After choosing the $z_4$-coordinate as the inner space,
using eq.~\eqref{eq:criticalpoints}, we get 
$\nu_{(24)} = 2$, and 
$\nu_{(4)} = 2 $. 
Accordingly, we choose the basis forms, 
\begin{equation}
\hat{e}^{(24)}_1=
\hat{h}^{(24)}_1=\frac{1}{z_2 z_4}, \qquad 
\hat{e}^{(24)}_2=\hat{h}^{(24)}_2=1 \, ,
\end{equation}
and for the inner space,
\begin{equation}
\hat{e}^{(4)}_1=
\hat{h}^{(4)}_1=\frac{1}{z_4}, \qquad 
\hat{e}^{(4)}_2=\hat{h}^{(4)}_2=1 \, .
\end{equation}

\noindent
$\bullet$ {$\rm \bf Cut_{\{2,4\}}:$}
On this specific cut, we use the regularized 
$
u_{2,4} = z_1^{\rho_1} z_3^{\rho_3} \,
u(z_1,0,z_3,0)
$
to obtain the corresponding ${\hat \omega}_1$ and ${\hat \omega}_3$. 
After choosing the $z_3$-coordinate as the inner space,
using eq.~\eqref{eq:criticalpoints}, we get 
$\nu_{(13)} = 2$, and 
$\nu_{(3)} = 2 $. 
Accordingly we choose the basis forms, 
\begin{equation}
\hat{e}^{(13)}_1=
\hat{h}^{(13)}_1=\frac{1}{z_1 z_3}, \quad 
\hat{e}^{(13)}_2=\hat{h}^{(13)}_2=1 \, ,
\end{equation}
and for the inner space,
\begin{equation}
\hat{e}^{(3)}_1=
\hat{h}^{(3)}_1=\frac{1}{z_3}, \quad 
\hat{e}^{(3)}_2=\hat{h}^{(3)}_2=1 \, .
\end{equation}

Now, with the help of eq.~\eqref{eq:masterdeco} and using eq.~\eqref{eq:multivarIntNumb} for 
the computation the individual multivariate (here $2$-form) intersection numbers, we 
determine the coefficients $c_i$ in eq.~\eqref{eq:masslessbox:decomp}. 

\noindent
{\bf Example.} Let us illustrate the decomposition of
\begin{align}
    \!\!\!\!\begin{gathered} \includegraphics[width=0.05\textwidth,valign=c]{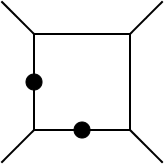} \end{gathered}  & =  \int_{\mathcal{C}} \frac{u \, d^4 \mathbf{z}  }{z_1^2 \, z_2^2 \, z_3 \, z_4} \ .
    \label{eq:box:example}
\end{align}
On the $\rm Cut_{\{1,3\}}$, we obtain:
\begin{align}
    \!\!\!\!\begin{gathered} \includegraphics[width=0.05\textwidth,valign=c]{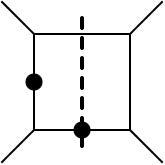} \end{gathered}  &
    =\int_{\mathcal{C}} u_{1,3} \, {\varphi}_{1,3} \, ,\quad \quad
    \varphi_{1,3}=\hat \varphi_{1,3}\, d {z_2} \wedge d {z_4} \ ,
\end{align}
where $\hat \varphi_{1,3} = \frac{\hat{\omega}_1  }{z_2^2 z_4}$.
On this specific cut we have:
\begin{equation}
\begin{gathered} \includegraphics[width=0.05\textwidth,valign=c]{box_dots_s_channel_cut.png} \end{gathered}= c_1 \begin{gathered} \includegraphics[width=0.05\textwidth,valign=c]{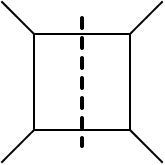} \end{gathered}+
c_2\, \begin{gathered} \includegraphics[width=0.05\textwidth,valign=c]{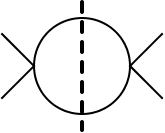} \end{gathered} \ ,
\end{equation}
with: 
\begin{align}
\label{eq:c1}
    &c_1 = \sum_{j=1}^{2} \langle \varphi_{1,3} | h_j^{(24)} \rangle \big({\mathbf C}_{(24)}^{-1}\big)_{j1} = \frac{(d-6) (d-5)}{s t}\, ,  \\
    &c_2 = \sum_{j=1}^{2} \langle \varphi_{1,3} | h_j^{(24)} \rangle \big({\mathbf C}_{(24)}^{-1}\big)_{j2} = -\frac{4 (d-5) (d-3)}{s^3 t} .\nonumber
\end{align}
On the $\rm Cut_{\{2,4\}}$, we obtain
\begin{align}
    \!\!\!\!\begin{gathered} \includegraphics[width=0.05\textwidth,valign=c]{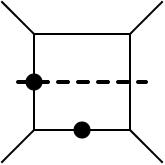} \end{gathered}  & = 
    \int_{\mathcal{C}} u_{2,4} \, \varphi_{2,4}\, , \quad \quad
    \varphi_{2,4}=\hat \varphi_{2,4}\, d {z_1} \wedge d {z_3},
\end{align}
where $\hat{\varphi}_{2,4}=\frac{\hat{\omega}_2}{z_1^2 z_3}$.
On this cut we have:
\begin{equation}
\begin{gathered} \includegraphics[width=0.05\textwidth,valign=c]{box_dots_t_channel_cut.png} \end{gathered}= c_1 \begin{gathered} \includegraphics[width=0.05\textwidth,valign=c]{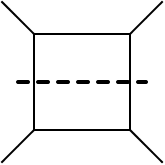} \end{gathered}+
c_3\, \begin{gathered} \includegraphics[width=0.04\textwidth,valign=c]{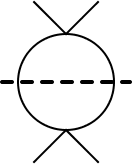} \end{gathered} \, ,
\end{equation}
where we find $c_1$ in agreement with eq.~\eqref{eq:c1} and
\begin{align}
    &c_3 = \sum_{j=1}^{2} \langle \varphi_{2,4} | h_j^{(13)} \rangle \big({\mathbf C}_{(13)}^{-1}\big)_{j2} = -\frac{4 (d-5) (d-3)}{s t^3}.
\end{align}
Finally, the integral of eq.~\eqref{eq:box:example} is decomposed in terms of MIs, as in eq.~\eqref{eq:masslessbox:decomp}, in agreement with the IBP decomposition.

\noindent
{\bf Differential Equation. } Let us consider the differential equation:
\begin{equation}
\partial_{s} \begin{gathered} \includegraphics[width=0.05\textwidth,valign=c]{box_no_label.png} \end{gathered}
=a_1  \begin{gathered} \includegraphics[width=0.05\textwidth,valign=c]{box_no_label.png} \end{gathered}
+a_2\, \begin{gathered} \includegraphics[width=0.05\textwidth,valign=c]{bubble_s_channel.png} \end{gathered}
+a_3\, \begin{gathered} \includegraphics[width=0.03\textwidth,valign=c]{bubble_t_channel.png} \end{gathered} \ ,
\end{equation}
where we restore the $s$-dependent prefactor:
\begin{equation}
 \begin{gathered} \includegraphics[width=0.05\textwidth,valign=c]{box_no_label.png} \end{gathered}=K \int_{\mathcal{C}}\frac{u \,  d^4 \mathbf{z}}{z_1 z_2 z_3 z_4} \, , \quad K=(-s t (s{+}t))^{2-\frac{d}{2}} \ .
\end{equation}
On the $\text{Cut}_{1,3}$ we obtain:
\begin{equation}
\partial_{s} \begin{gathered}
\includegraphics[width=0.05\textwidth,valign=c]{box_s_channel_cut.png}\end{gathered}
= K\int_{\mathcal{C}} u_{1,3}\, \varphi_{1,3} \, , \quad \varphi_{1,3}=\hat{\varphi}_{1,3} \, dz_2 \wedge dz_4
\end{equation}
with $\hat{\varphi}_{1,3}=\frac{f}{z_2 z_4}$ and $f=\frac{1}{K u} \frac{\partial(K u)}{\partial s}$.
On this cut we have:
\begin{equation}
\partial_{s} \begin{gathered}
\includegraphics[width=0.05\textwidth,valign=c]{box_s_channel_cut.png}\end{gathered}=a_1 \begin{gathered}
\includegraphics[width=0.05\textwidth,valign=c]{box_s_channel_cut.png}\end{gathered}+ a_2\, \begin{gathered}
\includegraphics[width=0.05\textwidth,valign=c]{bubble_s_channel_cut.png}\end{gathered}
\end{equation}
with
\begin{align}
\label{eq:a1}
    &a_1 = \sum_{j=1}^{2} \langle \varphi_{1,3} | h_j^{(24)} \rangle \big({\mathbf C}_{(24)}^{-1}\big)_{j1} = \frac{(d-6) t-2 s}{2 s (s+t)} \ ,  \\
    &a_2 = \sum_{j=1}^{2} \langle \varphi_{1,3} | h_j^{(24)} \rangle \big({\mathbf C}_{(24)}^{-1}\big)_{j2} = \frac{2 (d-3)}{s^2 (s+t)} \ .\nonumber
\end{align}
On the $\text{Cut}_{2,4}$ we have:
\begin{equation}
\partial_{s} \begin{gathered}
\includegraphics[width=0.05\textwidth,valign=c]{box_t_channel_cut.png}\end{gathered}=K \int_{\mathcal{C}} u_{2,4} \,  \varphi_{2,4}, \quad 
\varphi_{2,4}= \hat{\varphi}_{2,4} \, dz_1 \wedge dz_3
\end{equation}
with $\hat{\varphi}_{2,4}=\frac{f}{z_1 z_3}$.
On this specific cut we obtain:
\begin{equation}
\partial_{s} \begin{gathered}
\includegraphics[width=0.05\textwidth,valign=c]{box_t_channel_cut.png}\end{gathered}=a_1 \begin{gathered}
\includegraphics[width=0.05\textwidth,valign=c]{box_t_channel_cut.png}\end{gathered}
+ a_3\, \begin{gathered}
\includegraphics[width=0.04\textwidth,valign=c]{bubble_t_channel_cut.png}\end{gathered} \ ,
\end{equation}

where $a_1$ is in agreement with eq.~\eqref{eq:a1} and
\begin{align}
    &a_3 = \sum_{j=1}^{2} \langle \varphi_{2,4} | h_j^{(13)} \rangle \big({\mathbf C}_{(13)}^{-1}\big)_{j2} =-\frac{2 (d-3)}{s t (s+t)} \ .
\end{align}

\begin{figure}
\includegraphics[width=0.06\textwidth,valign=c]{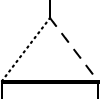}
\hspace*{0.2cm}
\includegraphics[width=0.07\textwidth,valign=c]{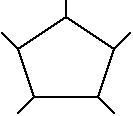}
\hspace*{0.2cm}
\includegraphics[width=0.07\textwidth,valign=c]{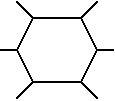}
\hspace*{0.2cm}
\includegraphics[width=0.09\textwidth,valign=c]{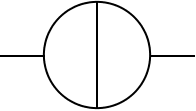}
\hspace*{0.1cm}
\includegraphics[width=0.07\textwidth,valign=c]{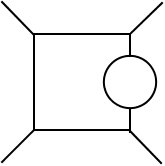}
\caption{Other examples of one- and two-loop integrals reduced to MIs with the technique proposed in this work.}
\label{fig:otherdiagrams}
\end{figure}

Let us finally remark that we have successfully applied the aforementioned algorithm to the complete decomposition of a few one- and two-loop integrals associated to   
the diagrams shown in Fig.~\ref{fig:otherdiagrams}, involving the evaluation of up to six-variable intersection numbers, and that the resulting 
expressions are in agreement with the IBP relations \cite{Smirnov:2014hma,Lee:2012cn,vonManteuffel:2012np,Maierhoefer:2017hyi}. 
Further examples are provided in the App. B.

\section{Conclusions}
\label{sec:conclusions}

Elaborating on the original proposal of \cite{Mastrolia:2018uzb} and on the wider studies \cite{Frellesvig:2019kgj,Mizera:2019gea}, we have shown that Feynman integrals can be expressed 
in terms of a complete basis of integrals, by making use of intersection numbers, 
which act as scalar products for the vector space of integrals, through the pairing of differential forms appearing in their integrands.
Let us notice that the final result of the recursion eq.~\eqref{eq:multivarIntNumb:IterativePsi} 
should not depend on the parametrization of the inner and outer space. 
Nevertheless, we observed that suitably chosen variable orderings may simplify and fasten the recursive procedure. This is a feature of the proposed algorithm that requires a dedicated study, which goes beyond the goal of the present work. 
Within Baikov representation, 
one-loop and multi-loop integrands have a similar structure, and therefore we expect that our decomposition algorithm can be applied to the case of integrals associated to more complex diagrams than the ones considered here, which we plan to investigate in the near future.

Scattering amplitudes are analytic functions, determined by their singularities. 
Intersection numbers, and their relation to Stokes' and Cauchy's residue theorems, embed what we believe is a clean role of analyticity in the amplitudes decomposition. 
We investigated the geometric origin of master integrals within the formalism of twisted (co)homology, where it was possible to relate them to the number of critical points and Euler characteristics in the connection to Morse/Picard--Lefschetz theory (as a special case of the 
Poincaré--Hopf index theorem).
Applications to Feynman integrals in representations other than Baikov will also constitute topics of future works.
The present study can be broadly applied in the context of theoretical particle physics, condensed matter and statistical mechanics, gravitational-wave physics, as well as mathematics.  \\

\noindent
{\bf Acknowledgments.} 
We would like to thank Janko B\"ohm, Thibault Damour, Maxim
Kontsevich, Stefano Laporta and Giovanni Ossola for useful discussions and comments. 
H.F., F.G., M.K.M, and P.M. wish to acknowledge the organizers and the participants of workshop {\it The Mathematics of Linear Relations between Feynman Integrals}, 18-22 March 2019, Mainz Institute for Theoretical Physics, Johannes Gutenberg University.
CloudVeneto is acknowledged for the use of computing and storage facilities. 
The work of F.G., M.K.M., and P.M. is supported by the Supporting TAlent in ReSearch at Padova University (UniPD STARS Grant 2017 ``Diagrammalgebra''). 
The work of H.F. is part of the HiProLoop project funded by the European Union’s Horizon 2020 research and innovation programme under the Marie Sk{\l}odowska-Curie grant agreement 747178.
Research at Perimeter Institute is supported in part by the Government of Canada through the Department of Innovation, Science and Economic Development Canada and by the Province of Ontario through the Ministry of Economic Development, Job Creation and Trade.
The figures were drawn with Jaxodraw~\cite{Binosi:2008ig} based on Axodraw~\cite{Axodraw:1994}.

\bibliography{references}

\appendix
\section{\label{sec:counting}Appendix A: Counting Master Integrals with Euler Characteristics, Morse Theory, and Lefschetz Thimbles}

Let us consider a single-valued $k$-form $\varphi_k$ and a multi-valued function $u(\mathbf{z})$ integrated over a $k$-real-dimensional submanifold $\mathcal{C}_k \subset X$ inside of some space $X$ of complex dimension $n$,
\begin{equation}\label{appendix-integral}
\int_{\mathcal{C}_k} u(\mathbf{z})\, \varphi_k(\mathbf{z}).
\end{equation}
If $u(\mathbf{z})$ regulates all boundaries of $\mathcal{C}_k$ then by Stokes' theorem:
\begin{equation}
0 = \int_{\mathcal{C}_{k}} d\left( u(\mathbf{z}) \varphi_{k-1} \right) = \int_{\mathcal{C}_k} u(\mathbf{z})\, \nabla_\omega \varphi_{k-1},
\end{equation}
where $\nabla_\omega \equiv d + \omega\wedge$ is a covariant derivative with a one-form $\omega \equiv d\log u(\mathbf{z})$. Thus adding terms of the form $\nabla_\omega \varphi_{k-1}$ to $\varphi_k$ does not change the value of the integral of eq.~\eqref{appendix-integral}. Similarly, we can impose that integrals over boundary terms of the form $\partial \mathcal{C}_{k+1}$ vanish:
\begin{equation}
0 = \int_{\partial\mathcal{C}_{k+1}}\!\!\! u(\mathbf{z})\, \varphi_k = \int_{\mathcal{C}_{k+1}}\!\!\! u(\mathbf{z})\, \nabla_{\omega} \varphi_k,
\end{equation}
which corresponds to $\nabla_\omega \varphi_k = 0$. These two requirements define a set of natural vector spaces for $k=0,1,\dots,2n$:
\begin{equation}\label{twisted-cohomology-definition}
H^k_{\omega} \equiv \{ k\text{-forms}\;\varphi_k \,|\, \nabla_\omega\varphi_k = 0 \} / \{ \nabla_\omega \varphi_{k-1} \},
\end{equation}
called \emph{twisted cohomology groups} \cite{aomoto2011theory}. Under some assumptions amounting to the fact that $u(\mathbf{z})$ regulates all boundaries of $X$, one can show that in fact $H^{n}_{\omega}$ is the only non-trivial space and all other $H^{k\neq n}_\omega$ vanish \cite{aomoto1975vanishing}. From now on we consider only such cases, even though Feynman integrals are known to sometimes violate these assumptions \cite{Lee:2013hzt,Frellesvig:2019kgj}.

One can also construct a dual vector space $(H^n_\omega)^\ast = H^{n}_{-\omega}$, with the same properties, given by a replacement $\omega \to -\omega$ in the above definition eq.~\eqref{twisted-cohomology-definition}. In this work we consider $\langle \varphi_L| \in H^n_{\omega}$ and $|\varphi_R\rangle \in H^{n}_{-\omega}$ and a scalar product $\langle \varphi_L | \varphi_R\rangle$ called the intersection number \cite{cho1995}%
\footnote{Similarly, eq.~\eqref{appendix-integral} is a scalar product $\langle \varphi_k | \mathcal{C}_k ]$ between $H^{k}_\omega \ni \langle\varphi_k|$ and the \emph{twisted homology group} $H_k^\omega \ni |\mathcal{C}_k]$, which is non-zero only for $k{=}n$. Since $|\mathcal{C}_n]$ is always constant in Feynman integral computations, $H^n_\omega$ can be also regarded as the vector space of Feynman integrals in a given family with the same $\omega$.}.

The Euler characteristic $\chi(X)$ of the space $X$ can be computed as an alternating sum of dimensions of $H^n_\omega$,
\begin{equation}
\chi(X) = \sum_{k=0}^{2n} (-1)^{k} \dim H^{k}_\omega.
\end{equation}
Since all $H^{k\neq n}_\omega$ vanish, we find that the dimension of $H^n_\omega$, and hence also the number $\masters$ of MIs is given by
\begin{equation}\label{Euler-characteristic-MIs}
\nu = (-1)^n \chi(X).
\end{equation}
Thus $\masters$ can be computed using one of the many ways of evaluating the topological invariant $\chi(X)$. We review a few of them below. Since $X = \mathbb{CP}^n {-} \mathcal{P}_\omega$, where $\mathcal{P}_\omega \equiv \{ \text{set of poles of }\omega\}$, we can simplify the above relation to
\begin{equation}\label{Euler-characteristic-complement}
\nu = (-1)^n\left( n{+}1  - \chi(\mathcal{P}_\omega)\right),
\end{equation}
where we used the fact that $\chi(\mathbb{CP}^n) = n{+}1$ and the inclusion-exclusion principle for Euler characteristics. The computation thus amounts to evaluating the Euler characteristic $\chi(\mathcal{P}_\omega)$ of the projective variety $\mathcal{P}_\omega$, see \cite{Aluffi:2008sy,Marcolli:2008vr,Bitoun:2017nre} for related approaches.

Let us introduce a simple function $u(\mathbf{z})$ that will serve as an instructive example in the remainder of this appendix:
\begin{equation}\label{appendix-u-example}
u(z) = \left( (z^2 {-} s^2)(z^2 {-} \rho^2) \right)^\gamma,
\end{equation}
which arises physically from the maximal cut of a two-loop non-planar triangle diagram \cite{Mastrolia:2018uzb} and gives rise to Appell $F_1$ functions with some constants $s,\rho,\gamma$. Computing $\omega = d\log u(z)$ gives straightforwardly $\mathcal{P}_\omega = \{\pm\rho,\pm s,\infty\}$, and hence $X=\mathbb{CP}^1 {-}\mathcal{P}_\omega$ is a one-dimensional space parametrized by an inhomogeneous coordinate $z$. The point at infinity is removed from $X$ since $\Res_{z=\infty}(\omega) \neq 0$. Since the Euler characteristic of $5$ distinct points is simply $\chi(\mathcal{P}_\omega)=5$, using eq.~\eqref{Euler-characteristic-complement} we find:
\begin{equation}\label{MIs-counting}
\nu = (-1)^1 \left(2 - 5 \right) = 3,
\end{equation}
which is the correct number of MIs in this case \cite{Mastrolia:2018uzb}.

Let us now consider a real-valued function $h(\mathbf{z}) \equiv \Re (\log u(\mathbf{z}))$, called a \emph{Morse function}, which assigns a ``height'' to every point $\mathbf{z} \in X$. Special role in this construction is played by \emph{critical points} $\mathbf{z}^\ast$ of $h(\mathbf{z})$ defined by $d h(\mathbf{z}^\ast) = 0$. Using Cauchy--Riemann equations it is straightforward to show that this condition is the same as $\omega = \sum_{i=1}^{n} \hat{\omega}_i dz_i = 0$ and thus the critical point equations read
\begin{equation}
\hat{\omega}_i = \partial_{z_i} \!\log u(\mathbf{z}^\ast) = 0, \qquad i=1,\dots,n.
\end{equation}
We assume that all critical points are isolated and non-degenerate. To each of them the Morse function assigns a pair of flows, labelled by a sign $\pm$ and parametrized by an auxiliary ``time'' variable $\tau$,
\begin{equation}
\frac{d z_i}{d\tau} = \mp \partial_{\overbar{z}_i} h(\mathbf{z}), \quad \frac{d\overbar{z}_i}{d\tau} = \mp \partial_{z_i} h(\mathbf{z}), \qquad i=1,\dots,n.
\end{equation}
In the $-$ case we have $dh(\mathbf{z})/d\tau < 0$ and hence it corresponds to a downward flow from the $\alpha$-th critical point $\mathbf{z}^\ast_{(\alpha)}$, which defines a submanifold of $X$ called a \emph{Lefschetz thimble} (or a path of steepest descent) $\mathcal{J}_\alpha$ with some real dimension $\lambda_\alpha$. Similarly, the $+$ case defines an upward flow, which generates a path of steepest ascent $\mathcal{K}_\alpha$ through the critical point $\mathbf{z}^\ast_{(\alpha)}$, with real dimension $2n{-}\lambda_\alpha$. Here $\lambda_\alpha$ is the number of unique negative directions extending from the $\alpha$-th critical point, called its \emph{Morse index}.

One of the key results in complex Morse theory (often called \emph{Picard--Lefschetz} theory) is that the Euler characteristic can be expressed as \cite{milnor2016morse}:
\begin{equation}\label{Euler-characteristic-Morse}
\chi(X) = \sum_{\lambda=0}^{2n} (-1)^{\lambda}\, \mathrm{M}_\lambda,
\end{equation}
where $\mathrm{M}_\lambda$ is the number of critical points with the Morse index equal to $\lambda$. Since $u(\mathbf{z})$ is a holomorphic function, near each $\mathbf{z}_{(\alpha)}^\ast$ we can pick local coordinates $\mathbf{w}_{(\alpha)}$ (with the critical point at $\mathbf{w}_{(\alpha)}{=}\mathbf{0}$) such that the Morse function admits an expansion:
\begin{equation}
h(\mathbf{w}_{(\alpha)}) = h(\mathbf{0}) + \Re\sum_{j=1}^{n} (w_{(\alpha),j})^2 + \dots.
\end{equation}
Treating $X$ as a real manifold with coordinates $\mathbf{w}_{(\alpha)} = \mathbf{x}_{(\alpha)} + i \mathbf{y}_{(\alpha)}$ we find
\begin{equation}
h(\mathbf{w}_{(\alpha)}) = h(\mathbf{0}) +  \sum_{j=1}^{n} (x_{(\alpha),j})^2 - \sum_{j=1}^{n} (y_{(\alpha),j})^2 + \dots
\end{equation}
and hence every critical point has a shape of a saddle with exactly $n$ upward and $n$ downward directions, or the Morse index $\lambda_\alpha = n$. This means that only $\mathrm{M}_{n}$ is non-vanishing and hence using eqs.~\eqref{Euler-characteristic-MIs} and \eqref{Euler-characteristic-Morse} we find \cite{aomoto1975vanishing}:
\begin{equation}\label{critical-point-counting}
\masters = \{ \text{number of solutions of } \omega{=}0\}.
\end{equation}
In the context of Feynman integrals these arguments were first given in \cite{Lee:2013hzt}.
The critical points can be also used to compute asymptotic behavior of intersection numbers \cite{Mizera:2017rqa}.

Let us mention that Lefschetz thimbles are integration contours along which eq.~\eqref{appendix-integral} converges the most rapidly for $k{=}n$, and thus the set $\{\mathcal{J}_\alpha\}_{\alpha=1}^{n}$ can be used a basis of integration cycles%
\footnote{Likewise, the paths of steepest ascent of $h(\mathbf{z})$, $\mathcal{K}_\alpha$ are integration cycles along which the dual integral $\int_{\mathcal{K}_\alpha}\!\! u(\mathbf{z})^{-1}\, \varphi_n$ converges the most rapidly and $\{\mathcal{K}_\alpha\}_{\alpha=1}^{n}$ can be used a basis of $H_n^{-\omega}$.}. For explicit examples of projecting cycles onto such bases using homological intersection numbers see App.~A of \cite{Mizera:2019gea}.

\begin{figure}[H]
\centering
	\includegraphics[scale=1]{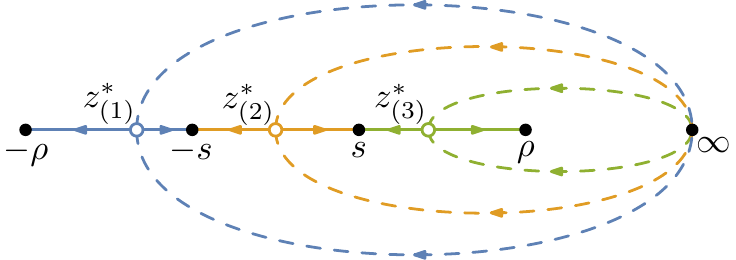}
	\caption{\label{fig:Morse-Smale-complex}Morse--Smale complex associated to the Morse function $h(z)=\Re (\log u(z))$ with eq.~\eqref{appendix-u-example} and $\rho{>}s{>}0$, $\gamma{>}0$.
	The set of filled dots corresponds to $\mathcal{P}_\omega = \{\pm \rho, \pm s,\infty \}$ removed from $X$. Empty dots at $z_{(\alpha)}^\ast$ represent critical points of the Morse function, with paths of steepest descent $\mathcal{J}_\alpha$ (solid lines) and ascent $\mathcal{K}_\alpha$ (dashed lines) extending from them. They give a triangulation of $X = \mathbb{CP}^1 {-} \mathcal{P}_\omega$. The arrows indicate the direction of the flow towards lower values of $h(z)$.}
\end{figure}

In the example at hand, eq.~\eqref{appendix-u-example} gives $\masters{=}3$ solutions of the critical point equations,
\begin{equation}
z^\ast = 0,\, \pm \sqrt{\frac{s^2 + \rho^2}{2}},
\end{equation}
in agreement with eq.~\eqref{MIs-counting}. The form of Lefschetz thimbles depends on the values of $s,\rho,\gamma$ and here we choose $\rho{>}s{>}0$ and $\gamma{>}0$ as a concrete example. With this choice each $\mathcal{J}_{\alpha=1,2,3}$ has to have endpoints on $z\in\{\pm \rho, \pm s\}$ since this is where $h(z)$ decays to $-\infty$, while $\mathcal{K}_{\alpha=1,2,3}$ can only have endpoints on $z=\infty$ as it is the only place where $h(z) \to +\infty$. This alone fixes the shape of the paths of steepest descent and ascent uniquely up to contour deformations. We illustrate them in Fig.~\ref{fig:Morse-Smale-complex}.

The critical points together with paths of steepest of descent and ascent triangulate the manifold $X$ into what is known as a Morse--Smale complex. Denoting the number of $q$-dimensional elements of this complex by $b_q$ (called the Betti number) we have
\begin{equation}
\chi(X) = \sum_{q=0}^{2n} (-1)^{q}\, b_q.
\end{equation}
For example, in Fig.~\ref{fig:Morse-Smale-complex} we can count $3$ vertices (the filled dots are not a part of $X$), $12$ edges (ignoring orientations), and $6$ faces. Together with eq.~\eqref{Euler-characteristic-MIs} this gives us yet another way of computing the number of MIs:
\begin{equation}
\masters = (-1)^{1} \left( 3 - 12 + 6 \right) = 3.
\end{equation}
For more background on Morse theory, see, e.g., \cite{milnor2016morse,Witten:2010cx} and in the context of twisted geometries \cite{aomoto1975vanishing,aomoto2011theory,Mizera:2017rqa,Mizera:2019gea}.
\section{\label{sec:massless-box_bubbleinsertion}Appendix B: A two-loop example}
\begin{figure}[H]
    \centering
    \includegraphics[width=0.12\textwidth]{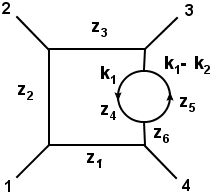}
    \caption{Massless box with a self-energy insertion diagram \\
    ($p_{i}$ with $p_{i}^{2}=0$, for $i=1,2,3,4$). The invariants are $s=(p_{1}+p_{2})^2$ and $t=(p_{2}+p_{3})^2$.}
    \label{fig:massless-box_bubbleinsertion}
\end{figure}

We consider the massless box with a self-energy insertion diagram at two loops, shown in  Fig.~\ref{fig:massless-box_bubbleinsertion}.
Within the \lbl BR~\cite{Frellesvig:2017aai}, $u$ reads:
\begin{align}
& u(\mathbf{z}) = \mathcal{B}_1^{\frac{2-d}{2}} \mathcal{B}_2^{\frac{d-3}{2}} \mathcal{B}_3^{\frac{d-5}{2}} \, ,
\end{align}
where
\begin{align}
\mathcal{B}_1 &= z_6 \, , \quad 
\mathcal{B}_2 = 2 (z_5{+}z_6) z_4 - z_4^2 - (z_5{-}z_6)^2 \, ,\nonumber\\
\mathcal{B}_3 &= t^2 z_1^2 + s^2 z_2^2 - 2 t z_1 ((2 s {+} t) z_3 {+} s (t {-} z_2 {-} z_6)) \\ 
& \;\;\; - 2 s z_2 (s t {-} t z_3 {+} (s {+} 2 t) z_6) + (t z_3 {+} s (z_6 {-} t))^2.\nonumber
\end{align}
For all the possible $63 \, (=2^6{-}1)$ sectors, using eq.~\eqref{eq:criticalpoints} on the corresponding cut, we determine the number $N_{sector}$ of MIs. The non-zero cases are: $N_{1,2,3,4,5}=1$, $N_{1,3,4,5}=1$ and $N_{2,4,5}=1$, giving $3$ MIs. We choose them as:
\begin{equation}
J_1=\begin{gathered} \includegraphics[width=0.05\textwidth,valign=c]{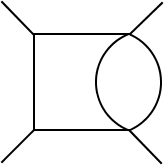} \end{gathered} \, , \quad
J_2=\begin{gathered} \includegraphics[width=0.05\textwidth,valign=c]{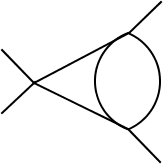} \end{gathered} \, , \quad
J_3=\begin{gathered} \includegraphics[width=0.03\textwidth,valign=c]{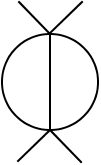} \end{gathered} \, .
\end{equation}
Any integral $I$ of the form \eqref{eq:def:I:basic}, with $u$ given in \eqref{eq:def:u:box}, and $\varphi$ defined in \eqref{eq:def:phi:FeynInt} (with $n=6$), can be decomposed as,
\begin{equation}
\begin{gathered} \includegraphics[width=0.05\textwidth,valign=c]{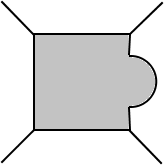} \end{gathered}= c_1 \begin{gathered} \includegraphics[width=0.05\textwidth,valign=c]{bubble_box.png} \end{gathered}+ c_2 \begin{gathered} \includegraphics[width=0.05\textwidth,valign=c]{triangle_bubble.png} \end{gathered}+ c_3\, \begin{gathered} \includegraphics[width=0.03\textwidth,valign=c]{sunrise.png} \end{gathered} \ . 
\label{eq:bubblebox:decomp}
\end{equation}
We use the set of spanning cuts ($\text{Cut}_{\{1,3,4,5\}}$, $\text{Cut}_{\{2,4,5\}}$) to obtain the full decomposition.\\
\noindent
$\bullet$ {$\rm \bf Cut_{\{1,3,4,5\}} : $}
On this specific cut, we use the regularized $u_{1,3,4,5}=z_2^{\rho_2} \, u(0,z_2,0,0,0,z_6)$ to obtain the corresponding $\hat{\omega}_2$ and $\hat{\omega}_6$. After choosing the $z_2$-coordinate for the inner space, using eq. \eqref{eq:criticalpoints}, we get $\nu_{(62)}=2$, and $\nu_{(2)}=2$. We choose the basis forms as:
\begin{equation}
\hat{e}^{(62)}_{1}=\hat{h}^{(62)}_{1}=\frac{1}{z_2} \, , \qquad
\hat{e}^{(62)}_{2}=\hat{h}^{(62)}_{2}=1 \, ,
\end{equation}
and
\begin{equation}
\hat{e}^{(2)}_{1}=\hat{h}^{(2)}_{1}=\frac{1}{z_2} \, , \qquad
\hat{e}^{(2)}_{2}=\hat{h}^{(2)}_{2}=1 \, .
\end{equation}
$\bullet$ {$\rm \bf Cut_{\{2,4,5\}} : $}
On this cut, we use the regularized $u_{2,4,5}=z_1^{\rho_1}z_3^{\rho_3}u(z_1,0,z_3,0,0,z_6)$
to obtain the corresponding $\hat\omega_1$, $\hat\omega_3$ and $\hat\omega_6$. Using eq. \eqref{eq:criticalpoints} we get $\nu_{(631)}=2$, $\nu_{(31)}=2$ and $\nu_{(1)}=2$. The basis forms are chosen as:
\begin{align}
\hat{e}^{(631)}_{1}&=\hat{h}^{(631)}_{1}=\frac{1}{z_1 z_3}, \qquad
\hat{e}^{(631)}_{2}=\hat{h}^{(631)}_{2}=1 \, ,
\nonumber \\
\hat{e}^{(31)}_{1}&=\hat{h}^{(31)}_{1}=z_1, \qquad
\hat{e}^{(31)}_{2}=\hat{h}^{(31)}_{2}=1,
\end{align}
and
\begin{equation}
\hat{e}^{(1)}_{1}=\hat{h}^{(1)}_{1}=z_1, \quad
\hat{e}^{(1)}_{2}=\hat{h}^{(1)}_{2}=1.
\end{equation}
Now, with the help of eq.~\eqref{eq:masterdeco} and using \eqref{eq:multivarIntNumb} for 
the computation the individual multi-variate (here 2 and 3-forms) intersection numbers, we 
determine the coefficients $c_i$ in \eqref{eq:bubblebox:decomp}.\\
\noindent
\textbf{Example. } Let us consider the decomposition of 
\begin{equation}
\begin{gathered} \includegraphics[width=0.05\textwidth,valign=c]{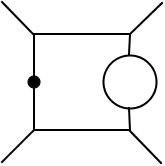} \end{gathered}=\int_{\mathcal{C}} \frac{u \,  d^6 \mathbf{z}}{z_1 z_2^2 z_3 z_4 z_5 z_6^2}.
\end{equation}
On the $\text{Cut}_{1,3,4,5} $, we obtain:
\begin{eqnarray}
\begin{gathered} \includegraphics[width=0.05\textwidth,valign=c]{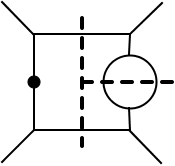} \end{gathered}
&=&
\int_{\mathcal{C}} u_{1,3,4,5} \, \varphi_{1,3,4,5} \, , \nonumber \\
\varphi_{1,3,4,5}&=&\hat{\varphi}_{1,3,4,5} \, dz_2 \wedge dz_6
\end{eqnarray}
where $\hat{\varphi}_{1,3,4,5}=\frac{\hat{\omega}_2}{z_2 z_6^2}$. 
On this specific cut, we have:
\begin{equation}
\begin{gathered} \includegraphics[width=0.05\textwidth,valign=c]{box_bubbleinsertion_4cut.png} \end{gathered}=c_1 \begin{gathered} \includegraphics[width=0.05\textwidth,valign=c]{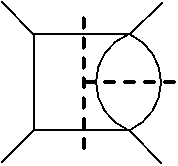} \end{gathered}+c_2
\begin{gathered} \includegraphics[width=0.05\textwidth,valign=c]{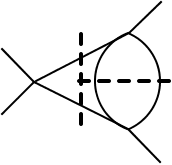} \end{gathered} \, ,
\end{equation}
with:
\begin{align}
c_1 & = \sum_{j=1}^{2} \langle \varphi_{1,3,4,5} | h_j^{(62)} \rangle \big({\mathbf C}_{(62)}^{-1}\big)_{j1} \nonumber \\
& = \frac{-3 (3d{-}16) (3d{-}14) (2 s{+}t)}{2 (d{-}6) s t^3} \, ,\\
c_2 & = \sum_{j=1}^{2} \langle \varphi_{1,3,4,5} | h_j^{(62)} \rangle \big({\mathbf C}_{(62)}^{-1}\big)_{j2}\\
& = \frac{-3 (3d{-}16) (3d{-}14) (3d{-}10) (2 d s{-}10 s{-}t)}{4 (d{-}6) (d{-}5) (d{-}4) s^2 t^3} \, .\nonumber
\end{align}
On the $\text{Cut}_{2,4,5} $, we obtain:
\begin{eqnarray}
\begin{gathered} \includegraphics[width=0.05\textwidth,valign=c]{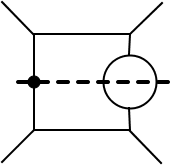} \end{gathered}
&=&\int_{\mathcal{C}} u_{2,4,5} \, \varphi_{2,4,5} \, , \nonumber \\
\varphi_{2,4,5}&=&\hat{\varphi}_{2,4,5} \, dz_1 \wedge dz_3 \wedge dz_6 \, ,
\end{eqnarray}
where $\hat{\varphi}_{2,4,5}=\frac{\hat{\omega}_2}{z_1 z_3 z_6^2}$.
On this specific cut, we have:
\begin{equation}
\begin{gathered} \includegraphics[width=0.05\textwidth,valign=c]{box_bubbleinsertion_3cut.png} \end{gathered}=c_1 \begin{gathered} \includegraphics[width=0.05\textwidth,valign=c]{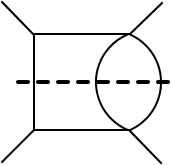} \end{gathered}+c_3 \begin{gathered}
\includegraphics[width=0.05\textwidth,valign=c]{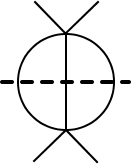} \ ,
\end{gathered}
\end{equation}
where $c_1$ is in agreement with the value found on the $\text{Cut}_{1,3,4,5}$, and 
\begin{align}
c_3 & = \sum_{j=1}^{2} \langle \varphi_{2,4,5} | h_j^{(631)} \rangle \big({\mathbf C}_{(631)}^{-1}\big)_{j2} \nonumber \\
& = \frac{3 (3d{-}16) (3d{-}14) (3d{-}10) (3d{-}8)}{2 (d{-}6)^2 (d{-}4) s t^4} \ .
\end{align}
Finally, we obtain a relation of the same type as in \eqref{eq:bubblebox:decomp}, in agreement with the IBP decomposition.

\end{document}